\documentclass[aps]{revtex4}
\usepackage{amssymb}

\usepackage{graphicx}
\usepackage{subfigure}

\begin{document}

\date{\today}
\title{A Fabry-Perot like two-photon interferometer for high-dimensional time-bin entanglement}
\author{Damien Stucki, Hugo Zbinden and Nicolas Gisin}
\affiliation{Group of Applied Physics, University of Geneva, 20 rue de l'\'{E}cole-de-M\'{e}decine, CH-1211 Geneva 4, Switzerland}

\begin{abstract}
We generate high-dimensional time-bin entanglement using a mode-locked laser and analyze it with a 2-photon Fabry-Perot interferometer. The dimension of the entangled state is limited only by the phase coherence between subsequent pulses and is practically infinite. In our experiment a pico-second mode-locked laser at 532 nm pumps a non-linear potassium niobate crystal to produce photon pairs by spontaneous parametric down-conversion ($SPDC$) at 810 and 1550 nm. 
\end{abstract}


\maketitle


\section{Introduction}

Entanglement is one of the most useful resources for quantum
information~\cite {tittel01}. Most entanglement based experiments
involved 2-level or eventually 3-level
systems~\cite{thew04A, molina04, langford04}. However over the last few years systems
with higher dimensions have received increasing attention for a variety of
reasons. The tolerance to noise of quantum key distribution can be
increased thanks to high-dimensional systems~\cite{cerf02}.
High-dimensional entanglement allows for the required
efficiency of detectors to close the detection loophole in EPR
experiments to be reduced~\cite{massar02}. Moreover, their properties
differ from classical ones more than 2-levels system and they have greater
robustness against noise~\cite{kaszlikowski00, collins02}.

High-dimensional systems can be obtained in two ways. Firstly, we
can get multi-photon (more than 2) entanglement by using high-order
parametric down-conversion~\cite{lamas01, howell2001,
weinfurter01}. Secondly, we can consider two-photon entanglement
in high-dimensional systems. This second approach has the
experimental advantage of higher coincidence count rates as you
have to create and detect only two photons. For example entanglement of
higher order angular momentum states of photons has been
demonstrated~\cite {mair01, vaziri01}. However, time-bins seem to
be the ideal scheme for higher dimensional entanglement. Indeed,
mode-locked laser can easily produce entangled states of almost
arbitrarily high dimensions. This has been shown using Michelson
interferometer~\cite{deriedmatten02, deriedmatten04}, i.e two-dimensional analyzers. Unfortunately, the
extension of this analysis to higher dimensions, using e.g.
interferometers with $n$ different paths, dramatically complicates
the experimental task. In this paper, we present an experimental
realization of a high-dimensional analysis using Fabry-Perot like
interferometers. We start with a theoretical description of our
analyzer before presenting the experiment and the results. 


\section{Theory}

A mode-locked laser is used to pump a non-linear crystal
in order to produce time-bin entangled photon pairs. We consider
a $D$-pulse train and assume a pair creation probability much
lower than $\frac{1}{D}$ per pulse to reduce the creation of two
pairs in a $D$-pulse train. When a photon pair is created in
time-bin $j$, the state is $\vert j,j \rangle$. As the time-bin in
which the photon pair is created is uncertain, the state after the
non-linear crystal is of the form:
\begin{equation}
\vert \psi_{crystal} \rangle=\sum_{j=1}^{D}c_{j}e^{i\phi _{j}}\vert j,j
\rangle  \label{EQ:stateAfterCrystal}
\end{equation}
where $c_j$ are the probability amplitudes and $\phi _{j}$ are the phase difference between successive pulses. For a mode-locked pump laser $c_{j}$ and $\phi _{j}$ are constant.

\begin{center}
\begin{figure}[tbp]
\includegraphics[width=12.5cm]{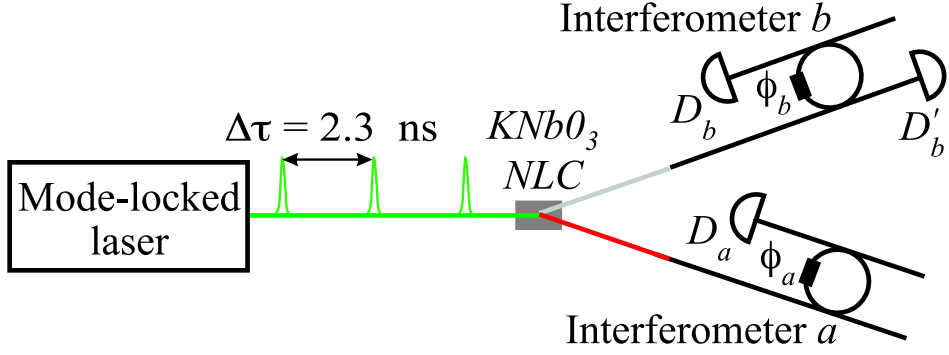}
\caption{Simplified scheme of the experiment. Pulses of a mode-locked laser
are sent through a $KNbO_3$ non-linear crystal (NLC) to produce photon pairs.
The two output modes are coupled into separated fibers and go through
interferometers with optical paths corresponding to the distance between two
successive pulses.}
\label{FIG:simpExperimentalScheme}
\end{figure}
\end{center}

\begin{center}
\begin{figure}[tbp]
\includegraphics[width=12.5cm]{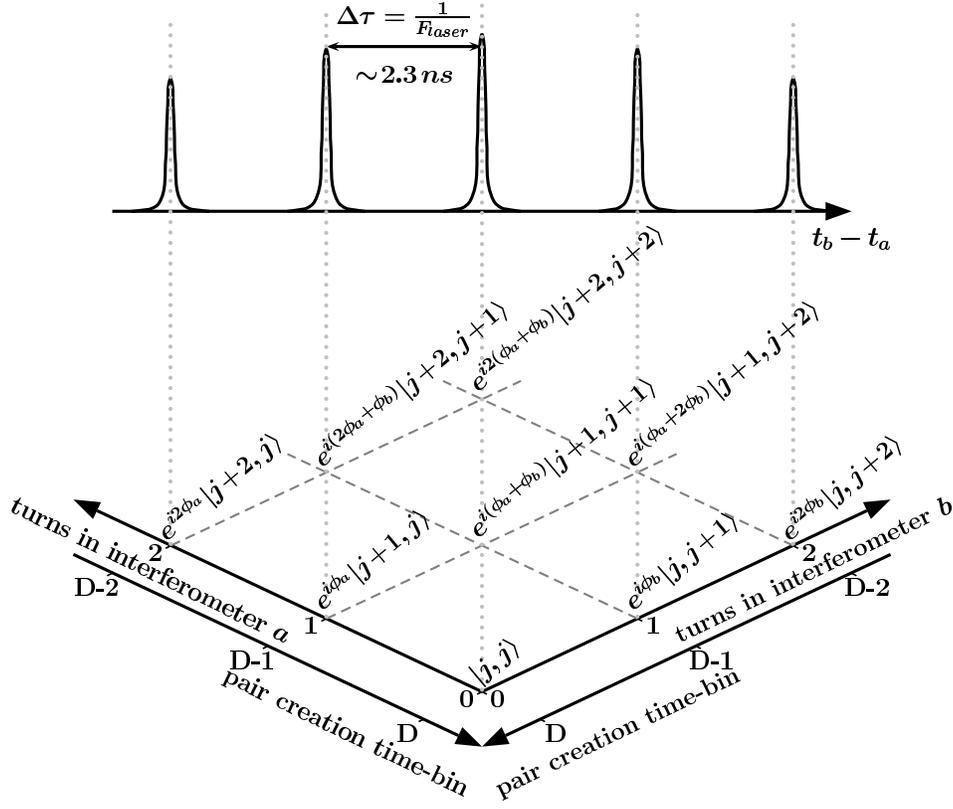} %
\caption{Coincidences as a function of the difference in arrival time for the two photons at detectors $D_{a}$ and $D_{b}$. These correspond to the sum of different interfering terms which are due to
the different possibilities for each possible arrival time difference.}
\label{FIG:coincidences}
\end{figure}
\end{center}

To analyze the high-dimensional time-bin entangled state we use
Fabry-Perot like interferometers (see Fig.~\ref{FIG:simpExperimentalScheme}). During one
turn through the interferometer a photon is delayed exactly by one time-bin $\Delta \tau =\frac{1}{f_{laser}}$ where $f_{laser}$ is the repetition frequency of the laser. Let us first consider the detectors $D_{a}$ and $D_{b}$.  After the interferometers where the two photons go to detectors $D_{a}$ and $D_{b}$, respectively, the state $\vert j,j \rangle$ evolves as follows~(the first passage through the interferometer to the detectors adds only a global
phase, which is not taken into account):
\newpage
\begin{eqnarray}
&&\qquad \qquad \qquad \qquad \qquad \qquad \vdots  \nonumber \\
&+&e^{i\phi _{a}}\vert j+1,j \rangle+e^{i(2\phi _{a}+\phi
_{b})}\vert j+2,j+1 \rangle+e^{i(3\phi _{a}+2\phi _{b})}\vert
j+3,j+2 \rangle+\dots
\nonumber \\
\vert j,j \rangle \rightarrow \quad &+&\vert j,j \rangle+e^{i(\phi
_{a}+\phi _{b})}\vert j+1,j+1 \rangle+e^{i2(\phi _{a}+\phi
_{b})}\vert j+2,j+2
\rangle+\dots  \label{EQ:stateAfterInterferometerJ} \\
\quad &+&e^{i\phi _{b}}\vert j,j+1 \rangle+e^{i(\phi _{a}+2\phi
_{b})}\vert j+1,j+2 \rangle+e^{i(2\phi _{a}+3\phi _{b})}\vert
j+2,j+3 \rangle+\dots
\nonumber \\
&&\qquad \qquad \qquad \qquad \qquad \qquad \vdots  \nonumber
\end{eqnarray}
where $\phi _{x}$ is the phase applied on the photons in mode $x=a$,
$b$ and can be considered constant for successive turns. 
\newline
The first row represents the situation when photon $a$ covers one more turn than photon $b$, second row represents photons $a$ and $b$ in the same time-bin and the third row represents the case with photon $b$ covering one more turn than photon $a$. 
\newline
The state of Eq.~(\ref{EQ:stateAfterCrystal}) then evolves, according to~(\ref{EQ:stateAfterInterferometerJ}), to:
\begin{eqnarray}
\vert \psi \rangle=
&\sum\limits_{n=0}^{D-1}\sum\limits_{j=1}^{D-n} c_{n,D} (e^{in\phi
_{a}}+e^{i((n+1)\phi _{a}+\phi _{b})}+\dots +e^{i(D\phi _{a}+(D-n)\phi
_{b})})\vert j+n,j \rangle&  \nonumber \\
&+\sum\limits_{m=1}^{D-1}\sum\limits_{j=1}^{D-m} \tilde{c}_{m,D} (e^{im\phi _{b}}+e^{i(\phi
_{a}+(m-1)\phi _{b})}+\dots +e^{i((D-m)\phi _{a}+D\phi _{b})})\vert j,j+m
\rangle&  \label{EQ:stateAfterInterferometer}
\end{eqnarray}
To analyze the system we measure the difference in the time of arrival of photons $a$ and $b$ at
detectors $D_{a}$ and $D_{b}$, respectively. For each value of $n$ or $m$ from Eq.~(\ref{EQ:stateAfterInterferometer}) there is a corresponding peak in the histogram of the time difference $\Delta t=t_{b}-t_{a}$. The central and highest peak corresponds to coincidences with $\Delta t=0$, while the first peak on its
right corresponds to $\Delta t=\Delta \tau $ and first peak on its left to $%
\Delta t=-\Delta \tau $ (see Fig.~\ref{FIG:coincidences}).

The relative height of these peaks, i.e. the probability of coincidences
between $D_{a}$ and $D_{b\text{ }}$ for different $\Delta t$, can be
calculated as follows (for $D\rightarrow \infty $ and without losses):
\begin{eqnarray}
P_{n=0} \equiv P_{0}  & \sim & (t_{1a} t_{1b} t_{2a} t_{2b})^2 \bigg| \frac{1}{1-r_{2a} r_{2b} r_{1a}  r_{1b} e^{i(\phi_{a}+\phi_{b})}} \bigg|^2 \nonumber \\
P_{n<0} & = & (r_{2a} r_{1a})^{2|n|} P_{0} \nonumber \\
P_{n>0} & = & (r_{2b} r_{1b})^{2n} P_{0} \label{EQ:coincidences}
\end{eqnarray}
where:

\begin{itemize}
\item  $P_{n}$ is the coincidences probability between $D_{a}$ and $D_{b}$
for the n-th peak in the arrival time difference histogram. By convention we
denote the peak corresponding to photons doing the same number of
turn in each interferometer by $n=0$, the first peak to the right is denoted by $n=1$%
, and so on and similarly with peaks on the left $n=-1$ for the first one and
so on.

\item  $t_{mx}$ ($r_{mx}$) is the transmission (reflection) amplitude of the coupler $m=1$ (2) for the first (second) one on the way of the photon in interferometer in mode $x=a$, $b$. By convention a "reflected" photon stays in the same fiber. One has to choose $t_{mx}$ $\ll r_{mx}$ in order to have a strong weighting of the terms that involve many turns in the interferometers, which are characteristic of high-dimensional entanglement.
\end{itemize}

For all $P_{n}$ terms the phase dependence is the same for all $n$ so the
different peaks of coincidences in the gate of detection on $D_{b}$ show synchronous
oscillations as a function of the sum of the phases in the interferometers, $\phi
_{a}+\phi _{b}$.
\newline
It's interesting to also calculate the probability to have coincidences
between the detectors $D_{a}$ and the third detector $D_{b}^{\prime }$ that we use as a control (see Fig.~\ref{FIG:simpExperimentalScheme}):
\begin{eqnarray}
P_{n=0}^{\prime} \equiv P_{0}^{\prime} & \sim & \bigg(\frac{t_{1a} t_{2a}}{r_{1b}} \bigg)^2 \bigg| -r_{1b}^{2} + \frac{t_{1b} t_{2b} r_{2a} r_{2b} r_{1a}  r_{1b} e^{i(\phi_{a}+\phi_{b})}}{1-r_{2a} r_{2b} r_{1a}  r_{1b} e^{i(\phi_{a}+\phi_{b})}} \bigg|^2 \nonumber \\
P_{n<0}^{\prime} & = & (r_{2a} r_{1a})^{2(|n|-1)} P_{0}^{\prime} \nonumber \\
P_{n=1}^{\prime} \equiv P_{1}^{\prime} & \sim & (t_{1a} t_{2a} t_{1b}^2 r_{2b})^2 \bigg| \frac{1}{1-r_{2a} r_{2b} r_{1a} r_{1b} e^{i(\phi_{a} + \phi_{b})}} \bigg|^2 \label{EQ:coincidencesPrime} \\
P_{n>0}^{\prime} & = & (r_{2b} r_{1b})^{2(n-1)} P_{1}^{\prime}\nonumber
\end{eqnarray}
where:

\begin{itemize}
\item  $P_{n}^{\prime }$ is the coincidence probability between $D_{a}$ and $%
D_{b}^{\prime }$ for the n-th peak in the histogram of arrival time difference.
As for $P_{n}$ we use the convention that $n=0$ for the case when photons in modes $a$ and $b$
go to the detectors with the same number of complete turns, $n=-1$ for the
first peak on its left and so on and $n=1$ for the first peak on its right and so on. Note that this
histogram is asymmetrical, as all peaks for $n>0$ are much smaller than those
for $n\leq 0,$ since $t_{1b}\ll r_{1b}$.
\end{itemize}

\begin{center}
\begin{figure}[tbp]
\includegraphics[width=10cm]{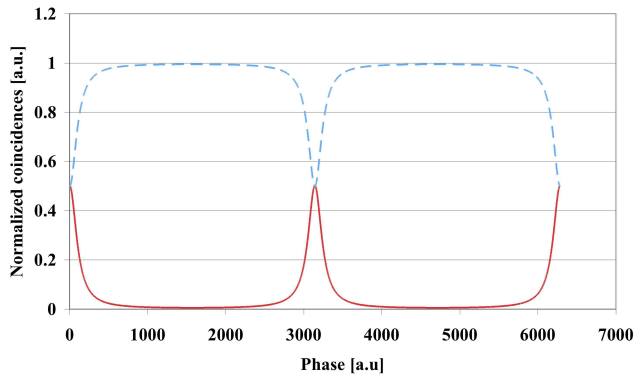}
\caption{Simulation of normalized coincidences as a function of the phase. The solid line corresponds to coincidences between $D_{a}$ and $D_{b}$ and the dashed line corresponds to coincidences between $D_{a}$ and $D_{b}^{\prime}$ ($D \rightarrow \infty$, $r_{mx} = \protect\sqrt{0.9}$).}
\label{FIG:simulation}
\end{figure}
\end{center}

The $P^{\prime }$ terms show the same behavior as the $P$ terms, however
we have a minimum of coincidences with detector $D_{b}^{\prime }$ when we
have a maximum of coincidences with detector $D_{b}$ as can be expected by conservation of energy. Indeed, the light has to go out of the interferometer $b$ by one of the two outputs if there is no losses (absorption) in the interferometers. This different behavior for the two terms is due to the minus sign in front of $r_{1b}^{2}$ in the formula of $P_{0}^{\prime }$ which is a consequence of the $\frac{\pi }{2}$ phase acquired when a photon is "transmitted", i.e. coupled. In Fig.~\ref{FIG:simulation} we can
easily verify that these probabilities sum up to unity in the case without
losses. We see that as a function of $\phi_{a}+\phi_{b}$, we do not obtain
a sinusoidal variation as we are used to in the case of qubits. The curves
remind us of the transmission through a Fabry-Perot interferometer,
which is a consequence of the high dimensionality of interferences, i.e. many interfering paths. The goal
is now to find this signature experimentally.

\section{Experiment}

\begin{figure}[tbp]
\includegraphics[width=15cm]{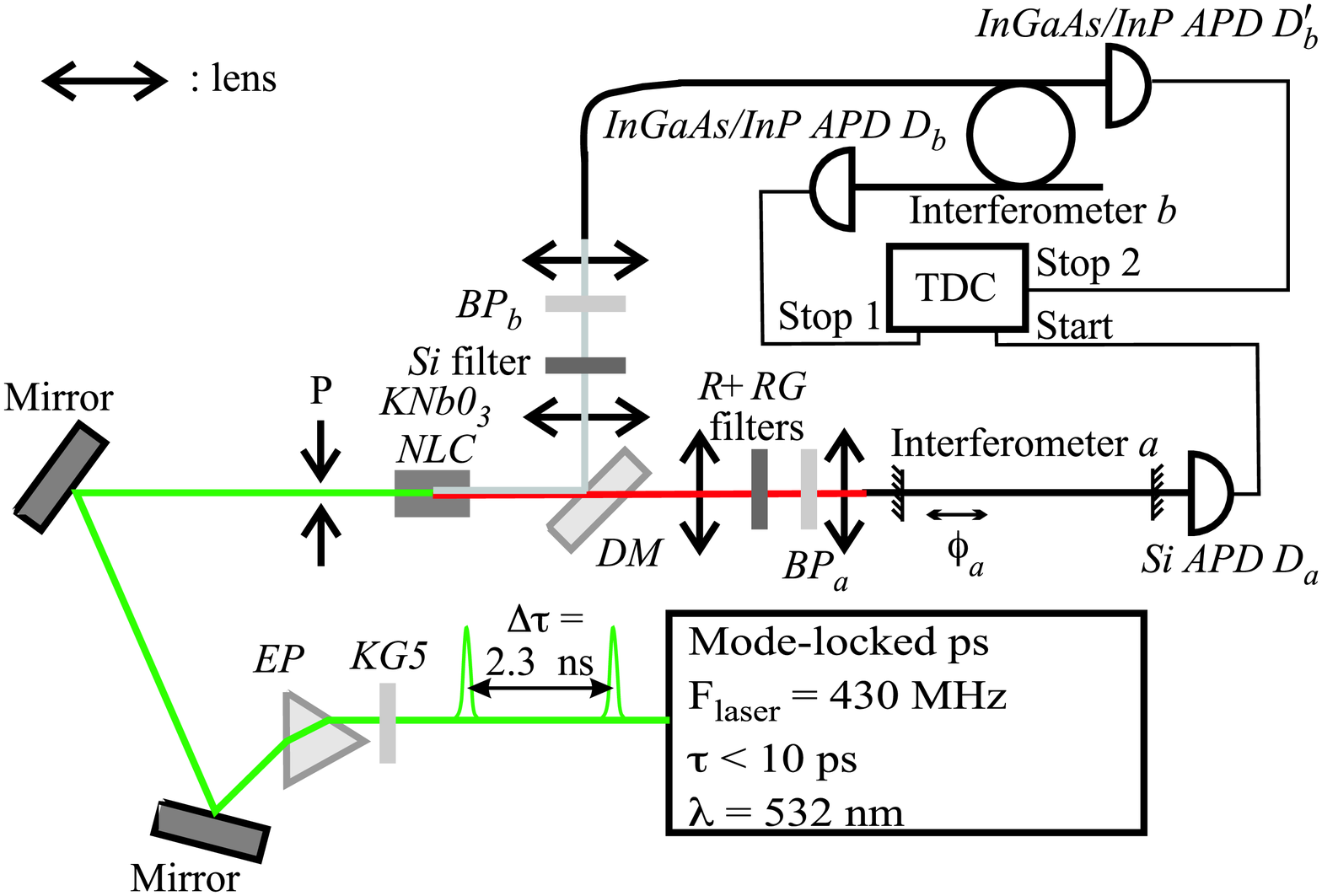}
\caption{Experimental scheme. The pulses of the mode-locked laser are sent through
a KG5 filter, an equilateral prism ($EP$) and a pinhole ($P$) to be
monochromatic. After that they go through a $KNbO_3$ non-linear crystal ($NLC $) and produce non-degenerate photon pairs at 810 and 1550 nm by type I parametric down-conversion. The photons at 810 nm are transmitted through a dichroic mirror ($DM$) and the
photons at 1550 nm are reflected. The photons in the 810 nm arm are filtered
by a reflector at 532 nm deposited on a RG-610 filter ($R+RG$ filters) and a bandpass
filter of 10 nm (FWHM) centered at 810 nm ($BP_a$). The photons in the 1550 nm arm
are filtered by a $Si$\ and a bandpass filters of 20 nm (FWHM) centered at 1550 nm ($BP_{b}$).
The photons are coupled into monomode (at their wavelengths) fibers and go
through the interferometers. Detection on the $Si$\ $APD$ $D_{a}$ triggers (not represented) the $InGaAs/InP$\ $APD's$ ($D_{b}$ and $D_{b}^{\prime}$) and starts the TDC.
Detections on the $InGaAs/InP$\ detectors stop the TDC.}
\label{FIG:experimentalScheme}
\end{figure}

A mode-locked, frequency-doubled Nd-laser (Time-Bandwidth GE-100, $\lambda
=532$ nm, FWHM$<10$ ps, $F_{laser}=430$ Mhz, $P_{mean}=30$ mW) is the heart
of our experiment (see Fig.~\ref{FIG:experimentalScheme}). A $f_{1}=200$ mm
achromatic doublet lens focalizes the light on a potassium niobate
non-linear crystal ($KNbO_{3}$, Castech, $\theta =23^{\circ }$, $\varphi
=0^{\circ }$) cut in order to obtain collinear signal and idler at 810 nm
and 1550 nm wavelengths by type I parametric down-conversion. A dichroic mirror is used to separate the two non-degenerated photons. In each output arm, a lens is firstly used to collimate the beam
while the second one focuses light into the monomode optical fiber at 810 nm
and 1550 nm respectively. As usual we also have to be very careful to filter out all
photons originating from the pump. First we remove the remaining photons at 1064 nm, using a KG5 filter, a dispersive equilateral prism and a diaphragm.
In order to remove the pump photons at 532 nm after the crystal, we put a
RG-610 filter coated with a dielectric mirror at 532 nm and a 10 nm (FWHM)
bandpass filter at 810 nm in arm $a$ and a combination of an AR coated silicon filter and a 20
nm (FWHM) bandpass filter at 1550 nm in arm $b$.
\newline
The interferometer $b$ is made of two $R/T=90/10$
couplers which are spliced together to the required length. An in-line fiber
polarization controller (Newport PolaRite F-POL-IL) is added in this loop.
The realization of interferometer $a$ is different to simplify alignment with the interferometer $b$ (see Fig.~\ref{FIG:experimentalScheme}). It is made from a monomode fiber at 810 nm of about 23.6 cm length with dielectric mirrors deposited on the
cleaved extremities with reflectivity and transmitivity $R/T=90/10$. The fiber is cut slightly shorter than it
normally should be and then it is stretched with a translation stage. A
piezoelectric actuator (PZT) allows us to then vary the length, i.e the
phase, by a few wavelengths. Both interferometers are enclosed in separated PI (proportional and integral parameters) temperature-regulated boxes.

Alignment of the interferometers is the first experimental
problem. The optical path lengths of interferometers $a$ and $b$ must be
the same to within the coherence length of the photon pairs, i.e 120 $\mu $m, 
as well as the cavity length of the pump laser, to within the coherence
length of the pump ($\sim 2$ mm in fiber). We use an auxiliary, bulk
Michelson interferometer where the path length difference is firstly adjusted to
interferometer $b$, using low coherence interferometry. We
then adjust the cavity length of the laser and of the interferometer $a$ to the
auxiliary Michelson interferometer.
\newline
The 810 nm photons are detected by a silicon ($Si$) single photon detector in passive
mode (EG\&G PQ-F830) and the 1550 nm photons are detected by a $InGaAs/InP$\ single
photon detectors (ID Quantique, id 200 SPDM) gated by the $Si$\ detector. The
gate width is 50 ns so we can detect 20 time-bins in each gate. The $Si$ detector output starts the Time-To-Digital Converter (TDC, ACAM AM-F1) and one stop is given by each output of the two $InGaAs/InP$\ detectors $D_{b}$ and $D_{b}^{\prime}$.

\section{Results}

\begin{center}
\begin{figure}[tbp]
\subfigure[]{\includegraphics[width=10cm]{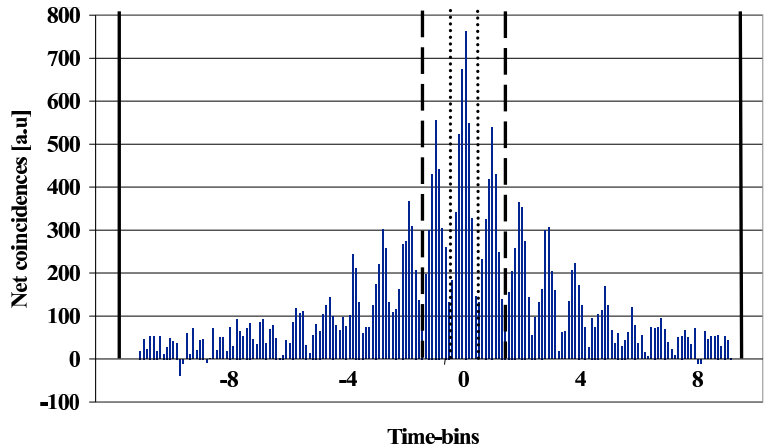}} \\
\subfigure[]{\includegraphics[width=10cm]{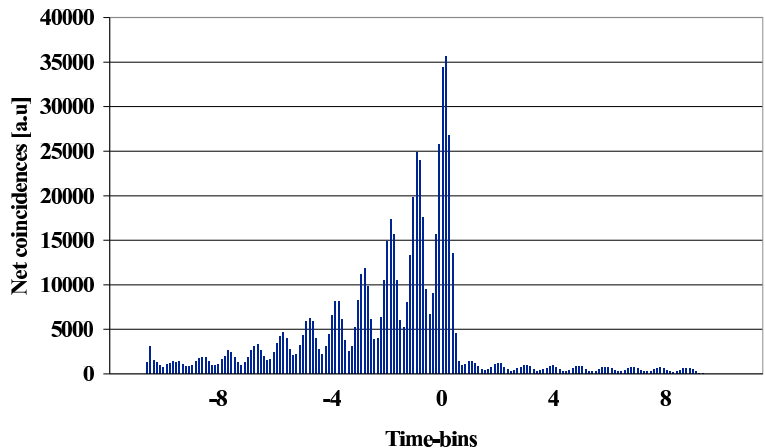}}
\caption{Net coincidences as a function of the difference of arrival times. The
histograms are the accumulation of a long term measurement. (a) Histogram
between detectors $D_{a}$ and $D_{b}$. The vertical lines represents the
different coincidence gates for Fig.~\ref{FIG:measureSynchro}, for the complete gate (solid line), the three
central peaks (dashed line) and the central peak (dotted line). (b)
Histogram between detectors $D_{a}$ and $D_{b}^{\prime}$. In the case (a)
the accumulated coincidences is lower than in case (b) because $T \ll R$.%
}
\label{FIG:histogram}
\end{figure}
\end{center}

\begin{figure}[tbp]
\includegraphics[width=10cm]{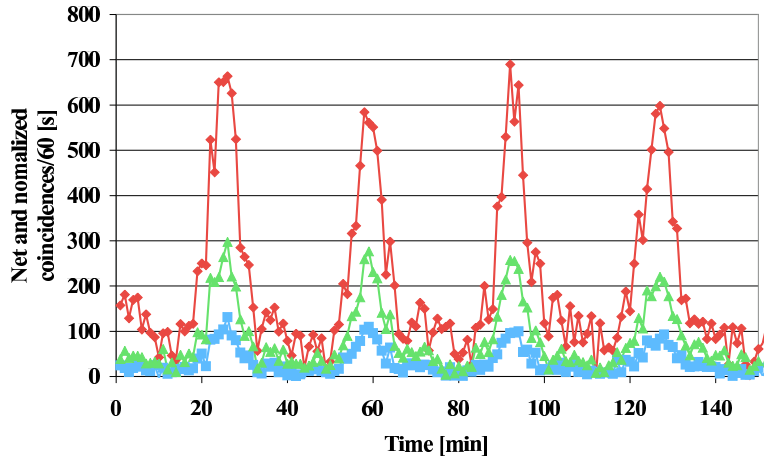}
\caption{Net and normalized coincidences between $Si$\ detector $D_{a}$ and $%
InGaAs/InP$ detector $D_{b}$ as a function of time while changing the phase (see text). We see interferences for the central peak ($\blacksquare$), the three central peaks ($\blacktriangle$) and the entire gate (about 20 peaks, $\blacklozenge$) (see Fig.~\ref{FIG:histogram}). The lines are only presented as a guide. }
\label{FIG:measureSynchro}
\end{figure}

Figs.~\ref{FIG:histogram} (a) and (b) show typical  histograms for the time
difference between a click from detector $D_{a}$-$D_{b}$ and  $D_{a}$-$D_{b}^{\prime }$, respectively, recorded with the  TDC. In the case
(a) the number of accumulated coincidences is much lower than in the case
(b), because most of the light is reflected on the first coupler of interferometer $b$ and goes
directly to the detector $D_{b}^{\prime }$. The vertical lines represent the
different time windows used in the measurements of Fig.~\ref{FIG:measureSynchro} . In all
measurements we record the number of coincidences
as a function of $\phi _{a}$, which is a function of the PZT voltage. For this purpose, we accumulate the number of coincidences, typically for 1 minute, then increase, step by step, the voltage on
the PZT. In order to minimize fluctuations due to varying pump power or
coupling of the down-converted photons into the fibers, we normalize all
coincidence rates with respect to the average single count rates of detector
$D_{a}$. We also subtract the noise of $InGaAs/InP$\ detectors, whereas the
dark-count of the $Si$\ detector ($D_{a}$) can be neglected. The noise of the $InGaAs/InP$\  detectors is due to the thermal dark count of the detector and it is of 15.6$\pm$0.1 Hz on $D_{b}$ and 17.6$\pm$0.1 Hz on $D_{b}^{\prime }$ for an efficiency of detection of about 16 and 18\%\ respectivly, for the entire gates of 50 ns and a gating frequency of about 4.6 kHz. 

This gating frequency correspond to the rate of detection on the $Si$ detector and it can be explained as follows. The repetition frequency of the laser is 430 MHz and the incident power on the crystal is approximately 17 mW and we have a probability of pair creation lower then 1\%, thus we have a pair creation frequency of $<$4.3 MHz. We can expect a global coupling factor of the order 10 \% between the crystal and the monomode fiber at 810 nm (including the losses through the filters). Therefore 430 kHz of photon at 810 nm are coupled into the fiber. With the losses of about -14 dB when the light goes through the interferometer $a$ and a detection efficiency for the $Si$ detector of the order of 40-50\% we can expect a detection frequency of few kHz. 

Firstly, we observe that the coincidences between $D_{a}$ and $D_{b}$ vary
synchronously for all different detection windows. In Fig.~\ref
{FIG:measureSynchro} we see the coincidences accumulated over the
entire gate of 50 ns, the three central peaks and only the central peak,
respectively. These three different coincidences sets oscillate synchronously
as expected.

\begin{figure}[tbp]
\includegraphics[width=10cm]{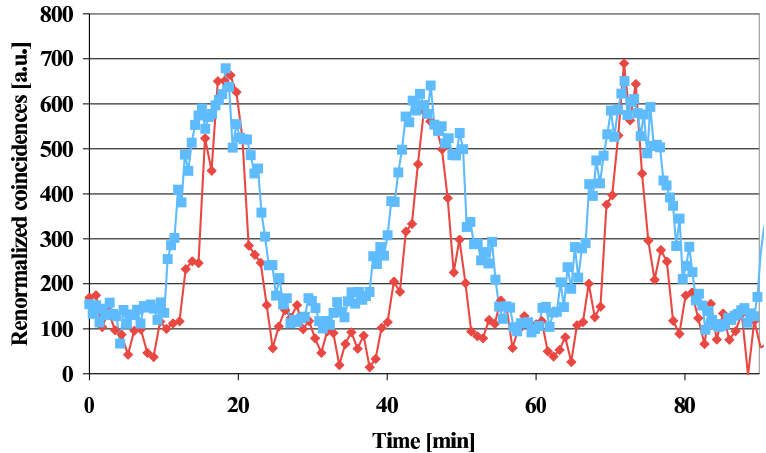}
\caption{Coincidences as a function of time (see text) with two different
fixation systems for the fiber in interferometer $a$ for the entire 50 ns $InGaAs/InP$ detectors gate. In the first case the fiber is squeezed within a metallic holder ($\blacksquare$) and in the second case
the fiber is glued on a metallic holder ($\blacklozenge$). The first measure is
renormalized to have the same amplitude and time dependance.}
\label{FIG:fixation}
\end{figure}

We notice that the peaks are considerably broader than what we would
expect for the ideal case according to Eq. (\ref{EQ:coincidences}) and depicted in
Fig.~\ref{FIG:simulation}. Of course the experiment is not perfect and we can improve our
theoretical model in order to take into account the following four experimental
limitations: a) there are
losses in the order of 5\% per round trip for interferometers $a$ and $b$. The
losses essentially reduce the contribution for the cases where both photons
make several roundtrips in the interferometers. 
b) High visibility interferences can only be achieved if the polarization states of the
interfering paths are identical. For this purpose, we inject in the interferometer, light from an external and pulsed laser, polarized in the same direction as the down-converted photons. We introduce then an auxiliary polarizer at the output of the interferometer. With the internal polarization controller, we now maximize the transmission of all the peaks, corresponding to zero, one, two and more roundtrips in the interferometer. Unfortunately, a perfect alignment is very difficult to achieve in practice and the remaining misalignment increases with the number of roundtrips. 
In the interferometer $a$ we don't have a polarization controller, we count on the
fact that in a short straight fiber, without stress induced birefringence,
the polarization is not altered, in principle. 
However, we have to pay attention to stress induced birefringence. In particular we realized that the peaks are narrower if we glue the fiber on a holder rather than fixing it by squeezing it in metallic holders (see Fig.~\ref{FIG:fixation}). 
c) We have to take into account that our light is not monochromatic
and hence the phase is not exactly the same for all wavelengths. 
d) Moreover, a slight fluctuation in temperature during the measurement can introduce
some phase noise in the order of $\pi /8$ per $0.01^{\circ }C.$ Again, this
phase noise adds up with each roundtrip and is hence more important for the
terms characterizing the high-order entanglement.

\begin{figure}[tbp]
\includegraphics[width=10cm]{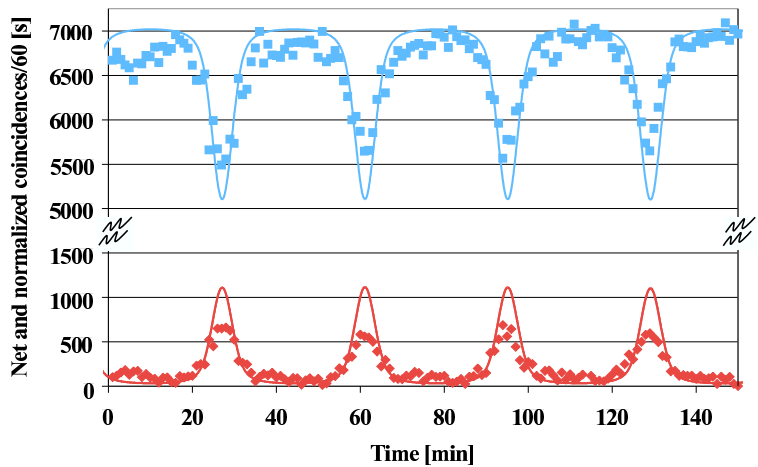}
\caption{Measurements and simulations of coincidences as a
function of time (see text) for the entire 50 ns $InGaAs/InP$ detectors gate. The measured coincidences between detector $D_{a}$-$D_{b}$ ($D_{a}$-$D_{b}^{\prime}$) are represented by $\blacklozenge$ ($\blacksquare $). For the simulations we consider spectrums of 5.4 nm at 810 nm and 20 nm at 1550 nm and gaussian fluctuations on the phase of about $\pi /8$\ FWHM corresponding to a length fluctuations of $\pm$25 nm of interferometer $a$.}
\label{FIG:comparison}
\end{figure}

We try to take into account these experimental limitations. We consider the losses as mentioned above and also the effective spectrums. The most limiting bandpass filter is the 20 nm at 1550 nm one and it correspond to a bandpass filter of 5.4 nm at 810 nm. Finally we introduce gaussian phase fluctuations of $\pi /8$\ (FWHM). We can qualitatively reproduce the measured curves (see Fig.~\ref{FIG:comparison}). Hence, we conclude that we have demonstrated the generation and the detection of high order entanglement.

\section{Conclusion}

We experimentally demonstrated high-order time-bin entanglement.
Whereas the creation of high-dimensional states is very convenient
with mode-locked lasers, we have to realize that the experimental
difficulties for the detection increase significantly with the
dimension of the Hilbert space. Therefore, despite some potential
advantages of high-dimensional entangled states discussed in the
introduction, these states tend to be of limited value for near future
practical applications.

\section{Acknowledgments}

This project was supported by European IST project RamboQ and the Swiss NCCR
quantum photonics. We would like to thank Michel Moret (GAP-B, University of
Geneva) for his technical support, Andreas Friedrich (IAP, University of Bern) for his help with dielectric coatings, Sofyan Iblisdir for helpful discussions and Rob T. Thew for the attentive reading of the paper. 

\bibliographystyle{apsrev}
\bibliography{biblio}

\end{document}